\documentclass[10pt,journal,compsoc]{IEEEtran}
\usepackage[utf8]{inputenc}

\usepackage{graphicx}
\usepackage{amsmath}
\usepackage[version=4]{mhchem}
\usepackage{siunitx}
\usepackage{longtable,tabularx}
\usepackage{comment}
\usepackage{listings}
\usepackage{amssymb}
\usepackage{color}
\usepackage[dvipsnames]{xcolor}
\usepackage{caption}
\usepackage{subcaption}
\usepackage{pdfpages}
\usepackage{float}
\usepackage{hyperref}

\begin{document}

\title{\Large{Weather of the Dorm Wi-Fi Ecosystem at the University of Colorado Boulder: Fall Semester 2019 to Spring Semester 2020 -- A Case Study of Wi-Fi including a Campus Response to the COVID-19 Perturbation}}

\markboth{Journal of \LaTeX\ Class Files,~Vol.~XX, No.~Z, Month~2021}%
{Shell \MakeLowercase{\textit{et al.}}: Bare Advanced Demo of IEEEtran.cls for IEEE Computer Society Journals}


\author{Jake Mcgrath {-Aeropace Engineering,}
        Armen Davis {-Applied Mathematics,} 
        Dr. James Curry -{-Applied Mathematics,}
        Orrie Gartner -{-Office of information Technology,} 
        Glenn Rodrigues -{-Office of information Technology,}
        Dr. Seth Spielman {-Office of Data Analytics,}
        Dr. Daniel Massey {-Department of Computer Science.}
        
Contact E-mail: james.h.curry@colorado.edu, Applied Mathematics, University of Colorado, Boulder 80309-0526.}



\date{November 20, 2020}

\IEEEtitleabstractindextext{

\begin{abstract}

"Growing use of network-enabled technology in Institutions of Higher Education (IHEs) among students, staff, and faculty means that there has been increasing demand to adapt technology platforms and tools that transform student learning strategies, faculty teaching, research modalities, as well as general operations.  In fact, many of the new modalities are a necessity for doing IHE business. In August 2019, our research team, at the University of Colorado Boulder, began collecting and analyzing data from the campus Wi-Fi network.  A goal of the research was to answer the question of "what passive sensing" of the IHE’s Wi-Fi might be able to tell you about the gross dynamics of the "Wi-Fi  weather" in the IHE ecosystem? Or more generally, what does anonymized data tell us about the dynamics in a IHE's ecosystem. Anonymized data were made available by the University of Colorado Boulder (CU) Office of Information Technology (OIT). Our goal was to understand the campus’ dynamical ecosystem as a reflection of its collected Wi-Fi data. Those data could then be used to develop forecast models and an understanding of the dynamics of the university ecosystem where the dynamics of Wi-Fi \textbf{ connected device count} could be used as a proxy for the ebb and flow of large scale, and small scale, behavior in the ecosystem. The analogy with weather prediction seemed appropriate and a viable strategy. Starting Fall 2019, data were collected in the 'observational phase' (data collection is ongoing). In the 'analysis phase,' briefly touched on here, we applied Singular Spectrum Analysis (SSA) "eigen decomposition," to deconstruct Wi-Fi data from dorms, the central campus dining cafeteria, the recreation center, and other buildings on campus. That analysis led to the identification of clusters of buildings that behaved similarly. Several campus buildings are dual use and they are placed in different clusters. Just as in the case of models of the weather, a final component of this research was forecasting. We found that weekly forecast of Wi-Fi behavior in the Fall 2019, was straight forward using SSA and seemed to present behavior of a low dimensional dynamical system. However, in Spring 2020, with the COVID-19 Perturbation, the campus ecosystem received a “shock” and data show that the campus changed very quickly. We found that as the campus moved to conduct remote learning, teaching, the closure of research labs, and the edict to work remotely, SSA forecasting techniques not trained on the Spring 2020, data after the shock, performed poorly. While SSA forecasting trained on a portion of the data did better, especially SSA R-forecasting. Finally, there is a short animation of the 'Wi-Fi Weather' for a typical Fall 2019 day on the CU Boulder campus, the full animation can be found at \href{https://youtu.be/jvygO73oNxk}{https://youtu.be/jvygO73oNxk}

\end{abstract}

\begin{IEEEkeywords}

Campus ecosystem, Wi-Fi Weather forecasting, Smart Communities, Streaming Data, Wi-Fi, Singular Spectrum Analysis, Covid-19 Perturbation.

\end{IEEEkeywords}}

\maketitle

\section{Introduction}

Examples of "smart and connected environments" are college and university campuses. Wi-Fi is also the most popular form of wireless technology adapted by Institutions of Higher Educations (IHEs) to provide network services to students, faculty and staff.  At the University of Colorado Boulder (CU Boulder) Wi-Fi officially launched around September 15\textsuperscript{th}, 1999 and is constantly evolving to yield faster data rates and better security. Wi-Fi currently operates on unlicensed frequencies in the 2.4GHz and 5GHz band, making it possible for any enterprise to deploy a Wi-Fi network that enables its stakeholders to be mobile and connect to both private and public networks. In addition, governing standards and certification bodies such as the IEEE and the Wi-Fi Alliance continue to improve Wi-Fi standards and ensure compatibility between Wi-Fi devices both current and anticipated. Through this research we have worked with the Boulder Campus Office of Informational Technology (OIT) to assess this profound perturbation in the educational ecosystem.

\smallskip

The starting point of this research was August 2019 at CU Boulder (or CUB) and has been focused on the campus Wi-Fi network usage. The larger research question was to what extent campus buildings have discernible WiFi characteristics and behaviors in their profiles from which we could generate information which would provide insights into the campus ecosystem, e.g. its "weather." This includes information on the impact of various large and small scale perturbations and disruptions of connected components of campus life. The research agenda is part of a multi-year program with an initial phase where we report on aspects of a "year in the life of a campus" from the perspective of Wi-Fi usage of a smart and connected community. Further, IHEs are possibly  at the epi-center of generations of the most tech savvy users and their presence on and off campuses is motivating rethinking issues around the emergent technology, cybersecurity, policy issues. Such  issues are ultimately being confronted by the larger society--IHEs are part of a societies cutting edge for innovation. Wi-Fi at CUB presents a rich  source of data and information, that could anticipate and model emergent patterns. Further, this view and the granularity of the data likely lends itself to construction of differential data-driven prediction (DDDP) modeling, that is differential equation models generated from data, that are akin to weather forecasting. 

CUB Wi-Fi data sets typically contain 57,109,890 entries and have a point density of 1472.42857143 points/day. It is reasonable to expect that with the increased use of network-enabled technology among IHEs, the data density should be expected to grow. In fact, on campus Wi-Fi deployment has resulted in increases in the spread of technology platforms and tools to transform the delivery of education. Many of these new learning modalities and technologies demand network connectivity for functionality. Further, because CUB is a research university, additional research devices are also demanding network access. So, developing and deploying mathematical and computational tools, methods and strategies for better understanding such data rich environment should lead to a better understanding of other data rich connected environments, e.g. smart communities. Research studies such as this one may lead to advance modelling of population centers and development of their infrastructure.

In Spring 2020, with the appearance of COVID-19 on the world stage this research took the opportunity to examine that pressing and emergent phenomena with the goal of both better understanding whether campus data might expose perturbations of normal and usual patterns of the connected campus as it tried to assess and also respond to changes in various campus environments such as dorms, study halls, recreational facilities, eateries and lecture halls. Our research had some success with the previously collected data in looking at the ecosystem up to the campus response to COVID-19.

\subsection{Overview}

This research has three parts: observation, analysis, and forecasting. In what follows, the project first gives a large scale view of the CU Boulder Wi-Fi and presents components that may be shared with most other IHEs. We present the raw data and make observations on the ebb and flow of campus activities using known Wi-Fi behavior as a surrogate for population and depicting this with a 'weather map.' The behavior includes identifying Wi-Fi hot spots as the campus population goes about its usual activities at the outset of the Fall 2019 academic year like attending classes, visiting the central campus cafeteria  or common study spaces, returning to dormitories, or the periodic leavings and returns to campus. The weather maps and raw data show such events, including Fall Break and major holidays. Unexpected, the data show  clusters of behavior. The data also shows the start of the spring semester and some of the consequences of the spring semester COVID-19 perturbation and its aftermath.

In the analysis we deconstruct the streaming data using Singular Spectrum Analysis (SSA). That analysis tool allows us to identify typical behaviors, their frequency content, and finer characteristics using the 'anonymized' surrogate Wi-Fi data. Anonymized data means no privacy rights were breached. The third and final component of this research presents a discussion of the results of using averaged Wi-Fi data in SSA forecast models.


 \subsubsection{Weather Forecasting Analogy}

At its core, this research work was motivated by the seminal work of \cite{Richardson_1965} L. F. Richardson  as well as  results from the early 1960’s of \cite{Savitzky_1964}, \cite{Vautard_1989}. In the case of Lewis Richardson, he, and others, started with the goal of weather forecasting \cite {Washington}. This then led to the collection of data and model development based on the data of which he was successful in building a modest forecast. In [10], the approach was to build least squares approximations and that led the researchers in [10] to a “differential prediction” model. Just as in [9], their approach was novel and led to other approximation strategies for constructing prediction. More recently, the work of Tu, Rowley, et al. [11] was formative for the proposed research.  References [10] and  [11] are like [9] in that they started with observation and moved to analysis and forecasting. Wi-Fi ecosystems are ripe for such an approach, one consistent with our goals. \cite {Vautard_1989} was also a transformational reference because it was among the first articles to meld dynamical systems with Singular Spectrum Analysis (SSA) of streaming data, and especially measures of the "dimension" of recurrent dynamical behavior which are used here. 

Observations give a picture of the ebb and flow of daily Wi-Fi traffic and then give a snapshot of the Wi-Fi ecosystem. Because the University of Colorado Boulder is a comprehensive research university, the focus in this article will be limited common areas that students are likely to navigate on a nearly daily basis dormitories, eateries, study spaces, etc. We do this by presenting anonymized data from traces of Wi-Fi activity, and specifically device counts. After presenting an initial trace, we focus on the "mean data" from dorms. This research officially started just before the launch of the Fall 2019 semester on "move in day." The collected data continues through the Fall semester break, the return of students from that break, and the second departure in March 2020 at the start of the COVID-19 ramp down. The ramp down was associated with the Spring 2020 semester break and the end of the academic calendar year. In a series of graphs, we present data traces of this evolution and identify important dates associated with official campus announcements. 

With the start of Spring 2020, we present data from the return of students to the campus, and initially of the normal return to campus modes of operations as noted in Fall 2019. In fact, it is striking that the unprocessed raw Wi-Fi data captures the normal semester start up until: the start of the COVID-19 perturbation. For CUB the Perturbation started around the beginning of March 2020. As with the earlier section on observation, we focus on the mean dormitory behavior and present the mean deviation of the number of connected Wi-Fi devices.  We also note that dorms that house primarily engineering students have a higher device count, while students in residential apartments exhibit device behavior similar to students living off campus. Also worth noting was that there was significant network traffic as students transitioned to remote education from off-campus sites, from around the country, after the official start of the enhanced COVID-19 Spring 2020 break.

So, with the available Wi-Fi sensor information 'Wi-Fi Weather' forecast for the campus ecosystem is possible and could be integrated into a larger multi-layered modeling effort that could be useful for developing a better understanding of the emerging dynamics in analogous complex ecosystems. And with even higher resolution data collection in the future, a more detailed local analysis of ebbs and flows should be possible. This would enable more accurate predictions of the IHE ecosystem. We end the the article with a summary of our findings, and concluding remarks as well as and a couple of future directions.

\section{SSA and Data Collection: SSA Analysis}

To further understand the ebb and flow of Wi-Fi usage across campus both mathematical and statistical  techniques of time series analysis were employed. The main analysis tool is Singular Spectrum Analysis (SSA), a well developed time series analysis technique used to decompose, in our context, streaming data signals into interpretable components and, ultimately, produce forecasts using an eigenvector and eigenvalue decomposition of the data matrices. We  refer to such a deta decomposition as generating sets of $Eigen Triples$. There are several excellent references on SSA including \cite{golyandina2001analysis}, and \cite{Elsner},  \cite{GolyZhig2013} see also \cite{Hassani_2018}.

\smallskip
A critical component of this analysis  is its use of the singular value decomposition (SVD) where the core idea is to decompose the signals from the data matrix into three components -- a trend, oscillations or seasonal variation, and a structure-less noise component \cite{GolyZhig2013}. Then, by "removing the underlying noise," a better understanding of the ebb and flow, e.g. the dynamical behavior of the system, can be gleaned and a "smooth forecasts" can be extracted. Data streams with prominent oscillatory behavior especially benefit from SSA techniques since such oscillations can then be examined separately from other patterns at play. With multidimensional time series, an analogous technique called Multi-channel SSA or MSSA can be used to achieve similar results \cite{Goly_filteringof}.

With SVD being the dominant analysis technique used in both SSA and MSSA, data analysis leads to a significant flexibility, and freedom, as we shall see, in subsequent sections where details are presented. However, and as expected, additional flexibility comes with added complexity in constraints on SSA, i.e. how one chooses to decompose the underlying signal, indeed, parameter selection, is more of an art and is guided by the given data streams.

As we proceeded, a potentially useful scientific metaphor that was driving our research, because we are collecting access-point data from across campus, was an analogue to modern-day weather forecasting. More specifically, since such streaming data were collected at specific locations in the campus ecosystem, and forecasts based on the spatio-temporal data was desired, we will view SSA as being analogous to the technique of empirical orthogonal functions (EOFs) used by \cite{Richardson_1965} L.F. Richardson and also E.N. Lorenz \cite{Lorenz_1969} in their quest for increasingly accurate weather predictions.

\subsection{SSA Embedding}
We start with a time series $\mathbb{X} = (x_1, x_2, ..., x_N)$ of length $N\geq 2$ with at least one nonzero data point. To apply SVD to a one-dimensional time series, a standard process referred to  as \textit{embedding} takes place where the data stream is converted to a Hankel data matrix. With $N$ data points sampled in time and with a set \textit{window length} $L$, we can generate the following data matrix:

\begin{equation}
    \centering
    X = [X_1:X_2:...:X_K] = 
    \begin{bmatrix}
    x_1 & x_2 & ... & x_K\\
    x_2 & x_3 & ... & x_{K+1}\\
    x_3 & x_4 & ... & x_{K+2}\\
    \vdots & \vdots & \ddots & \vdots \\
    x_L & x_{L+1} & ... & x_N
    \end{bmatrix}
    \label{eq:3.1}
\end{equation}

with each column being a \textit{lagged vector} of the form

\begin{equation}
    \centering
    X_i = [x_i,...,x_{i+L-1}]^T \thinspace , \thinspace 1\leq i\leq K \thinspace, \thinspace K = N-L+1 \label{eq:3.2}
\end{equation}

\smallskip
Note that the elements in matrix \ref{eq:3.1} are equal on the off-diagonals. Such matrices need not be  square, and are  referred to in the literature as a \textit{Hankel Matrix}. This is the matrix representation we will use in the singular value decomposition of our streaming data matrices. Also it is possible to return to the one-dimensional time series by tracing the top/right or left/bottom edges of the matrix in equation \ref{eq:3.1}. Choosing the value of $L$ is problem specific, and is constrained to be $2 \leq L \leq N/2$. Higher values of $L$ offer more eigen triples components in the decomposition, but at the cost of increased computational complexity and more difficulty in the "grouping stage" as discussed in \cite{GolyZhig2013}. 

\subsection{SSA - Singular Value Decomposition}

The core idea behind SSA is the singular value decomposition (SVD). Like its counterpart, principle component analysis, we aim to use SVD to decompose the data matrix in equation \ref{eq:3.1} into its rank-one components where the positive singular values are arranged in decreasing magnitude. We begin by factoring

\begin{equation}
    \centering
    X = U\Sigma V^T
    \label{eq:3.3}
\end{equation}

where the columns of $U$ are the eigenvectors of $XX^T$, the columns of $V$ are the eigenvectors of $X^TX$ and the entries of the diagonal matrix $\Sigma$ are the eigenvalues of $XX^T$ (the singular values of $X$). Since $X \in \mathbb{R} ^{L \times K}$, it can be shown that $U \in \mathbb{R} ^{L \times L}$, $\Sigma \in \mathbb{R} ^{L \times K}$ and $V \in \mathbb{R} ^{K \times K}$. If we set $d = max\{i, \Sigma_{ii} > 0\} = rank(X)$, then it is possible to decompose the data matrix $X$ into $d$ rank-one components

\begin{equation}
    \centering
    X = X_1 + X_2 + ... + X_d
    \label{eq:3.4}
\end{equation}

where $X_i = U_i \Sigma_{ii} V_i^T$. In most time series, $d = min\{K, L\}$, \cite{GolyZhig2013}. Such a component $X_i$ is called an \textit{Eigen Triple} or ET for short. \textbf{Importantly}, the singular values of $\Sigma$ are in descending order, with the largest being in $\Sigma_{1, 1}$. The columns of $U$ and $V$ must be arranged accordingly, but this is usually done automatically by most modern software.

\smallskip

\section{A snapshot of the CU Boulder Wi-Fi Ecosystem: 
Dorms}

\smallskip

In the remainder of this introduction, we provide a snapshot of five of the critical areas in the University of Colorado Boulder Wi-Fi environment.
\par 
In what follows, we present a table of some of the buildings where data were sampled as part of the CU building ecosystem. These are buildings where large numbers of Wi-Fi device were counted throughout typical days during the academic year, Fall 2019 and Spring 2020.

\subsection{Dormitories}

Below is a partial list of dormitories included in this study. The data included in the table includes building name, a tentative group of students by major or  or college, number of rooms and reported bed count. When developing the average Wi-Fi device count, these data were helpful in understanding the ebb and flow of average connected Wi-Fi  device count. We present some of the information that was available in the Fall 2019. 

\medskip

\begin{table}[h!]
\centering
\begin{tabular}{ | p {2 cm} | p {2 cm} | p {1 cm} | p { 1 cm} | }

\hline
\bf{Dorm} & \bf{Students} & \bf{Rooms} & \bf{Beds} \\
\hline
Aden & Double Majors & 56 & 113 \\
\hline
Bear Creek B & 2nd years + & 172 & 498 \\
\hline
Buckingham & Media& 110 & 205 \\
\hline
Crossman & Engin & 46 & 138 \\
\hline
Smith & Arts & 185 & 338 \\
\hline
Stearns (E,W) & Out of State & 418 & 855 \\
\hline
Will Vill North & Older Dorm & 278 & 500 \\
\hline
Will Vill East & New Dorm & 420 & 705 \\
\hline
\end{tabular}

\caption{Dormitories examined on the CUB campus}
\label{tab:dedvice_count}
\end{table}


The Center for Community (C4C) is the main campus dining hall on the CU Boulder campus. It offers a range of food options and cuisines. The facility opens early and closes late. Like the Norlin Library, C4C is also "multiple use" facility and hosts a range of meeting rooms, offices and the like.

\smallskip

The fact that these two facilities are multiple use implies they are facilities that house campus offices with multiple missions. For example, the C4C houses some offices of student services focused on how to access community resources, medical leave, health and safety referrals, etc. More importantly, when students take a break from campus, e.g. during November's Fall break, many of the service offices continue to be operational and data traces show a small number of connected Wi-Fi devices. The same low level of connectivity is also visible, typically over the weekends when building offices are closed. Observations like this allowed us to establish a baseline for device count as the campus moves into the Spring 2020 semester and the COVID-19 perturbation.

\smallskip

\section{Wi-Fi Weather Maps-"WFW"}

Previously, we discussed Wi-Fi density in terms of devices per bed, or more usefully, devices per person. While this is useful for understanding device behavior at the building level, it is also beneficial to imagine the campus as an ecosystem.  To do this, we introduce another form of Wi-Fi density: devices/area or "physical Wi-Fi density." Physical Wi-Fi density is a useful measure because it illustrates exactly where Wi-Fi traffic originates on campus and how the macro scale traffic moves throughout the day. An apt analogy is that each campus building can be thought of a "weather station" in a Wi-Fi sensor network. Then, a goal is to "track" the weather in the ecosystem over an extended period. (See \cite{Savitzky_1964} for another approach.)

\subsection{Analogy and Methods: Building the Weather Maps}

To create this illustration, we first sourced a map of the entire CU Boulder campus that we used as a background for our density data. Next, we obtained the pixel coordinates of every building on campus from the campus map. For any given time, say 10/21/2019 at 8:00AM, our algorithm obtains the number of Wi-Fi devices in each building, finds that building's boundaries on the map, then fills the building with one point for each device. Then we use a Gaussian Kernel Density Estimation (KDE) function to estimate the probability density function of Wi-Fi propagation across the map. The map is then broken into discrete hexagons that are colored from the results of our KDE. Finally, contour lines are calculated and overlaid onto the colored hexagons. This process is repeated for every time step, and can be used to create animated maps. 

The result of this process looks very similar to a weather map where each building acts like a "weather station" observing nearby devices and reporting their numbers back to a central server. Then, as the day progresses, we are able to see an analogue of "pressure" (Wi-Fi density) systems move across campus, storms develop in certain areas, and even parts of campus that have become devoid of Wi-Fi activity completely, see \cite{Richardson_1965}, \cite{Malone_1955} and \cite{Lorenz_1963}.

\subsection{The Physical Campus Layout}

\begin{figure}[htbp]
    \centering
    \includegraphics[width=0.45\textwidth]{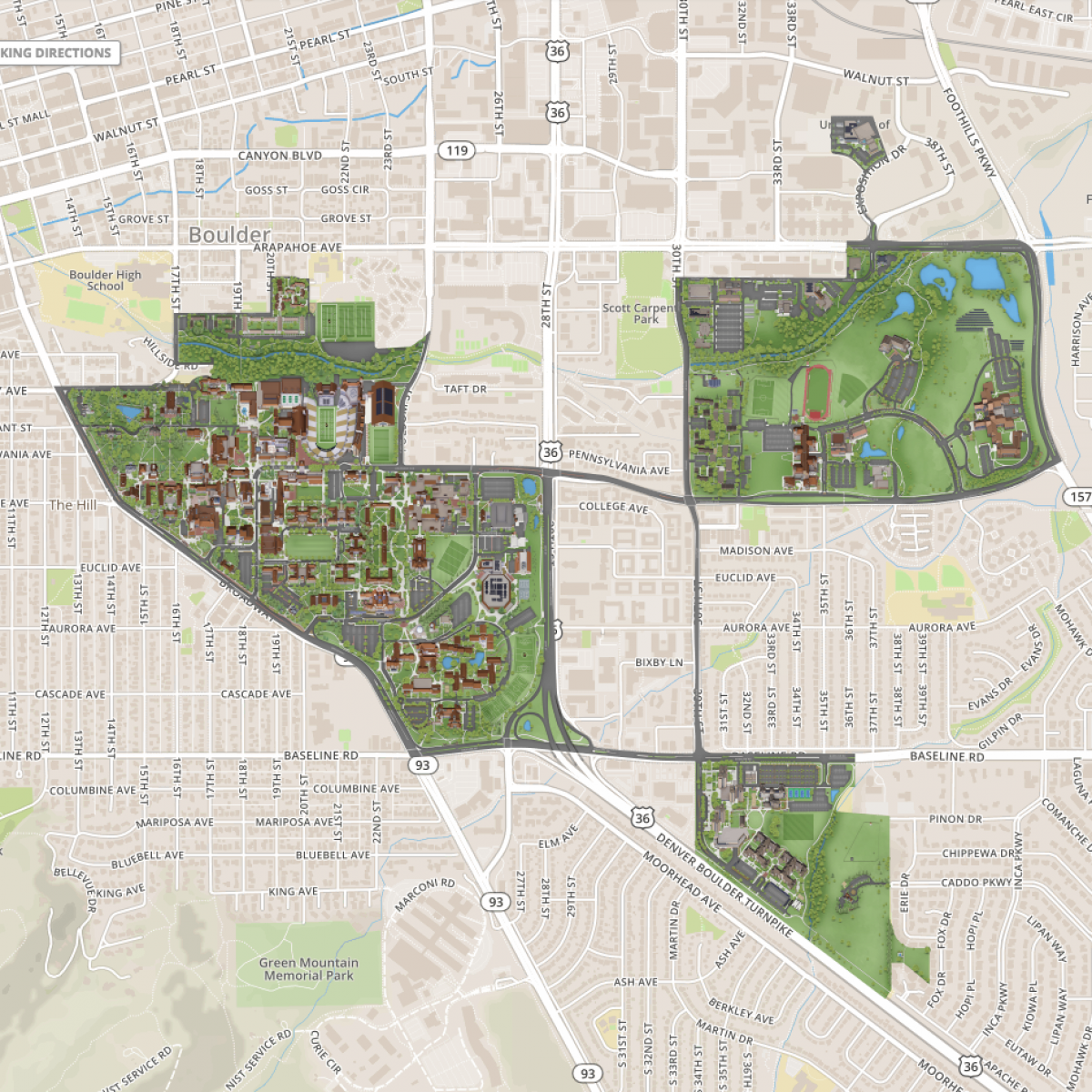}
   \caption{The CU Boulder Campus}
    \label{fig:blankmap}
\end{figure}

Wi-Fi dynamics suggest that the CU Boulder Campus is separated into three main regions outlined in figure \ref{fig:blankmap}. The largest, leftmost region is main campus and consists of scattered dormitories, learning spaces, eateries, and shared spaces like the recreation center and the library. The smaller, bottom most region southeast of the main campus is William's Village. It  consists of high capacity, high rise dorms, the Bear Creek Apartments, a dining hall, and a smaller recreation center. The final section in the upper right portion of the map is East Campus, which primarily contains  buildings specialized to a specific major like the aerospace building and some smaller eating spaces. 

\subsection{Ebb and Flow of Wi-Fi on Campus and an Animation}

The following snapshots come from a larger animation showcasing the change in Wi-Fi density across campus over a regular day in the Fall 2019 semester. An animation of the Wi-Fi weather for a typical Fall day on the CU Boulder campus can be found here: \href{https://youtu.be/jvygO73oNxk}{https://youtu.be/jvygO73oNxk}

In what follows, several discrete images are provided, first of the CU Boulder campus and subsequently of several activity hot spots that demonstrate the dynamics of Wi-Fi density on campus. All maps are from early October 2019 and show various Wi-Fi activity levels as the day progresses.

At 8:00AM, we observed the highest density of devices in William's Village and activity associated with dorms on Main Campus, including but not limited to the engineering dorms and also other dorms bordering Farrand Field. At this time of day there are comparatively no devices present on East Campus. This is understandable because many classes in this area start around 8:30AM, so we would not expect to see much activity.

Around noon, the density shifts away from William's Village and towards East and Main Campus. As previously stated, data suggests that many students from William's Village leave for the duration of the school day and focus most of their activities on the main campus. However, we also notice a shift away from dorms on Main Campus to educational buildings like Norlin Library, Duane Physics, Eaton Humanities, Cristol Chemistry, the Mathematics building, and the Engineering Center. The "pressure" map also indicates that there are two main hot spots on East Campus, the Aerospace Building and the Biotech building, which are two of the larger learning spaces in that area. 

At 3:00PM, we observe a shift to secondary and tertiary hot spots on main campus. The Wi-Fi density has begun to move away from educational spaces and back to some main campus living spaces such as Kittredge Commons, at the southern extreme of the central campus, and the Engineering Dorms, and to outdoor spaces like Farrand Field and Duane Field. Also, at this time, there is still a considerable concentration of students present in the Engineering Center and East Campus.

\begin{figure}[htbp]
    \centering
    \includegraphics[width=0.45\textwidth]{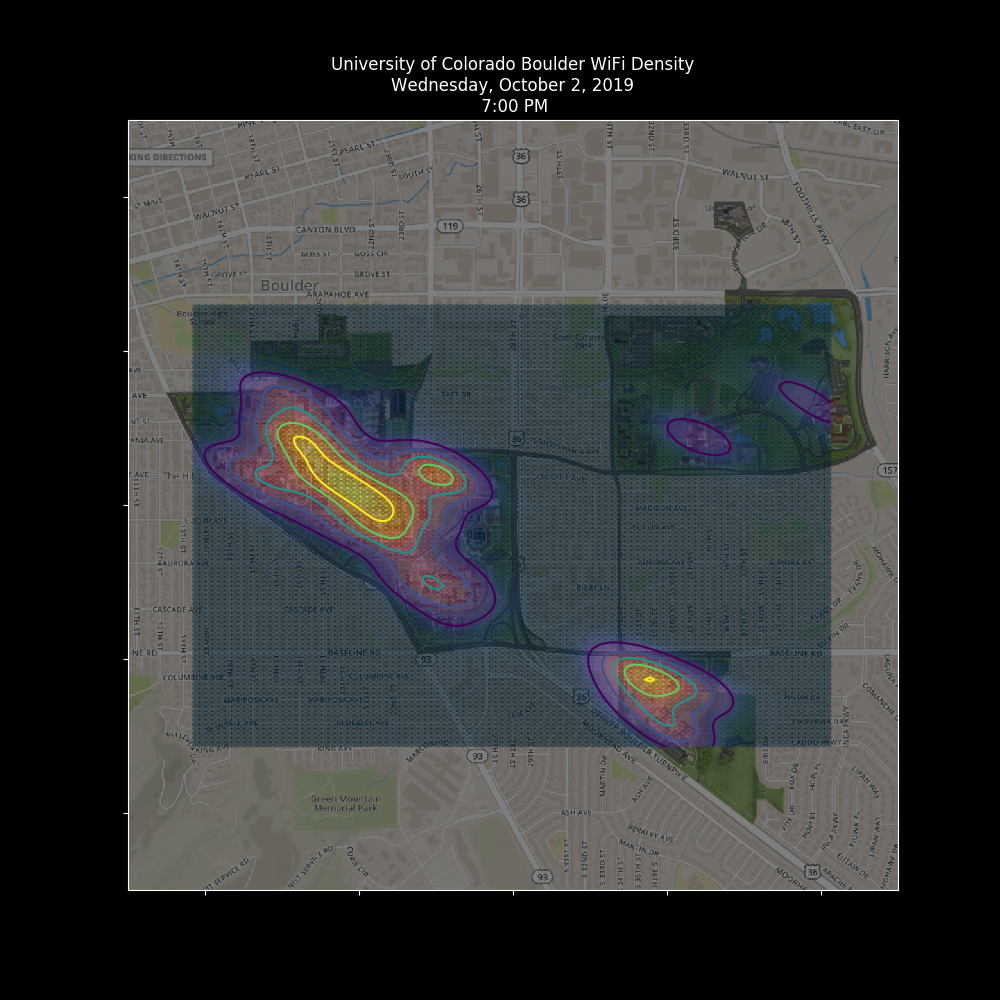}
    \caption{Wi-Fi Weather Map of CU Boulder Campus, 7:00 PM, Fall 2019}
    \label{fig:weather7pm}
\end{figure}

Finally, at 7:00PM, a shift of devices back to dorms as well as a shift in concentration to the main campus dining hall, C4C, the Center for Community. However, there are still notable concentrations of devices in the Engineering Center and East Campus at this time, but overall students seem to be settling in for the evening. 

\subsection{Interpreting the Hot Spots}

The brighter yellow portions of the weather maps are "hot spots," more concentrated areas of Wi-Fi devices, and thus more students. The shift of these hot spots during the day are significant because they show popular areas of campus and how the hive of devices migrates from place to place. It is reasonable to expect that this notion will become increasingly important in the Fall 2020 semester. With the rise of COVID-19, such high concentrations of devices could be indicative of transmission vectors for the virus. During a normal semester, these hot spots are a positive indication of a healthy campus; students are in class, enjoying time outside, spending time in the dining hall, and living together in unique communities. While we can't control where or how students move around the campus ecosystem, working to recognize and mitigate these hot spots as best as possible might be a strategy. In Fall 2020, a more diffuse weather map may well be a better indication of a healthy campus. 

\subsection{Granularity of Data leading to better predictions}

The maps showcased in this article were generated building by building. While this is satisfactory for understanding the campus ecosystem as a whole, finer detail may likely be required for more accurate predictions. Our data set supports this capability; we use spooled data on an access point level which we later aggregated to save computational power. Then, given an architect's blueprint of a given building as well as the placement of associated access points, we could generate maps of device movement on a building level. This has numerous practical applications which include, but are not limited to the following: 

\begin{itemize}
    \item Better planning and spacing of study tables in a given building. 
    \item An understanding of which places are used and when. 
    \item A more adaptable cleaning and sanitation schedule which limits worker contact and student disruption (particularly applicable for Fall 2020). 
    \item An understanding of how many devices are present in a given space and adapt Wi-Fi architecture accordingly.
    \item Understand how devices move through human designed spaces and optimize furniture arrangement. 
    \item Maximize floor space usage. 
    \item Security applications like alerts when devices are moving in buildings and shouldn't be. 
    \item Find network security vulnerabilities by visualizing where and when devices are connected to the network without human supervision (i.e vending machines that accept credit cards, Wi-Fi thermometers, lab equipment, workout equipment, digital picture frames, smart devices, etc.)
    \item Controlling the density of devices in a physical space by throttling network resources. 
    \item Limiting devices in a certain space to help reduce COVID-19 transmission vectors. 
\end{itemize}

Clearly, going forward with the Wi-Fi weather maps may provide additional utility or insights.

\section{Mean Dorm Behavior: Examining the Raw Data}
Going forward, we will also use the notion of a "mean dorm" in order to understand the relationship between the typical CU Boulder student and their Wi-Fi enabled devices. While device count does serve as a proxy representation of flow and movement of students on campus, device count over-represents the number of students in a given area. This is because there is a delay in the hand-off between Wi-Fi access points and device. The landscape of the classroom is quickly evolving and is becoming a more Wi-Fi connected environment. Further, it is no longer the case that individual students carry only one Wi-Fi enabled device to class. In fact, it has become increasingly common for students to tote a smartphone, laptop computer, and even a tablet (for taking notes). Furthermore, individual dormitories on campus seemingly have their own Wi-Fi fingerprints both in terms of physical and student makeup. As noted in Table 1, a CU Boulder dormitory can be as large as 850 beds or as small as 100 while some dorms even group students of similar majors to create learning communities. The variation in dorm composition and the fact that students often carry more than one device make it difficult to infer building-level behavior; these difficulties can be reduced by focusing on the average dorm, i.e. the "Mean Dorm."

\subsection{Methods}

While we have no way of knowing how many students are in a given classroom at a given time, we do know roughly how many students live in a dormitory. Each dorm has an available bed count, room count, and student grouping outlined in Table \ref{tab:dedvice_count}. We can use bed count and the knowledge that each student is given exactly one bed to normalize each dormitory for analysis and to determine roughly how many devices students carry with them on average. 

For each data set involving a dorm, we can divide the device count by the number of beds to obtain "devices per bed," or more usefully, "devices per person." We now have an understanding of how the number of devices and the device density (devices per person) changes throughout the day. We can use this information to establish an accurate range for the number of devices a student carries on a dorm by dorm basis, to normalize and directly compare two unlike dorms, and then create a fingerprint of the selected dorms behavior on average. 

\subsection{Average Device Count}

We expect that different groupings of students will have different devices. For example, the need for engineering students to have a powerful, separate laptop for CAD and simulations while an arts and sciences major could use just a tablet might skew the device counts of some dorms. The results of device count range from a normal week in Fall 2019 are outlined below in Table \ref{tab:density_counts}. 

\begin{table}[h]
\centering
\begin{tabular}{|l|l|l|l|}
\hline
Dorm & Max (D/P)& Min (D/P) & Diff(D/P) \\
\hline
Will Vill North & 1.30 & 0.88 & 0.42       \\
\hline
Sterns & 1.36 & 0.61 & 0.75       \\
\hline
Bear Creek Ap& 1.59 & 0.63 & 0.96       \\
\hline
Aden & 1.72 & 0.88 & 0.83       \\
\hline
Crossmam & 1.43 & 0.72 & 0.71       \\
\hline
Buckingham  & 1.20 & 0.66 & 0.54       \\
\hline
Smith & 1.29 & 0.70 & 0.59       \\
\hline
Will Vill East  & 1.34 & 0.51 & 0.83       \\
\hline
Mean                   & 1.36 & 0.81 & 0.55 \\     
\hline
\end{tabular}

\caption{The maximum, minimum, and average number of devices per person for selected dorms for CU Boulder campus.}
\label{tab:density_counts}
\end{table}

The minimum figure outlined in the table was often met very late at night or very early in the morning while the max was met around 8-9PM. On average, for every ten people in a dorm, between about eight and fourteen devices will be connected to the network. This information could be useful to those planning the Wi-Fi architecture so the network can be better equipped to handle student traffic. 

Overall, our hypothesis that more technologically demanding majors warranted more devices was supported. Aden Hall, concentrated with double majors, had the highest device density at 1.72 devices/person. Buckingham Hall, concentrated with media majors, had the lowest device density at 1.20 devices/person. The second highest density was found in Bear Creek Apartments, upperclassman housing, at 1.59 devices/person, which can likely be attributed to the fact that these students have larger living spaces and thus more room for Wi-Fi creature comforts like gaming consoles or smart devices. We similarly find the next highest density in Crossman Hall, an engineering dorm, and the second lowest density in Smith hall, an arts dorm. 

\subsection{Some Typical Usage Patterns}

Now that the notion of a mean dorm has been established, we can normalize and compare unlike dorms over time to find deviations from the average in day to day behavior. The graphs showcased in this section are information dense; the purple line represents the normalized device count (devices/person) for the dorm in question, the black line represents the mean dorm device count, the blue shaded areas showcase the standard deviation from the mean (darker is smaller deviation), and finally the red and green shapes across the bottom are the difference between the dorm's device normalized count and the mean normalized device count (green means the dorm has a higher normalized device count, red means a lower normalized device count). 

The snapshots in this subsection were taken from a standard week in Fall 2019 and is a good representation of standard behavior for the dorms. 

\begin{figure}[h!]
	\centering
    \includegraphics[width=0.45\textwidth]{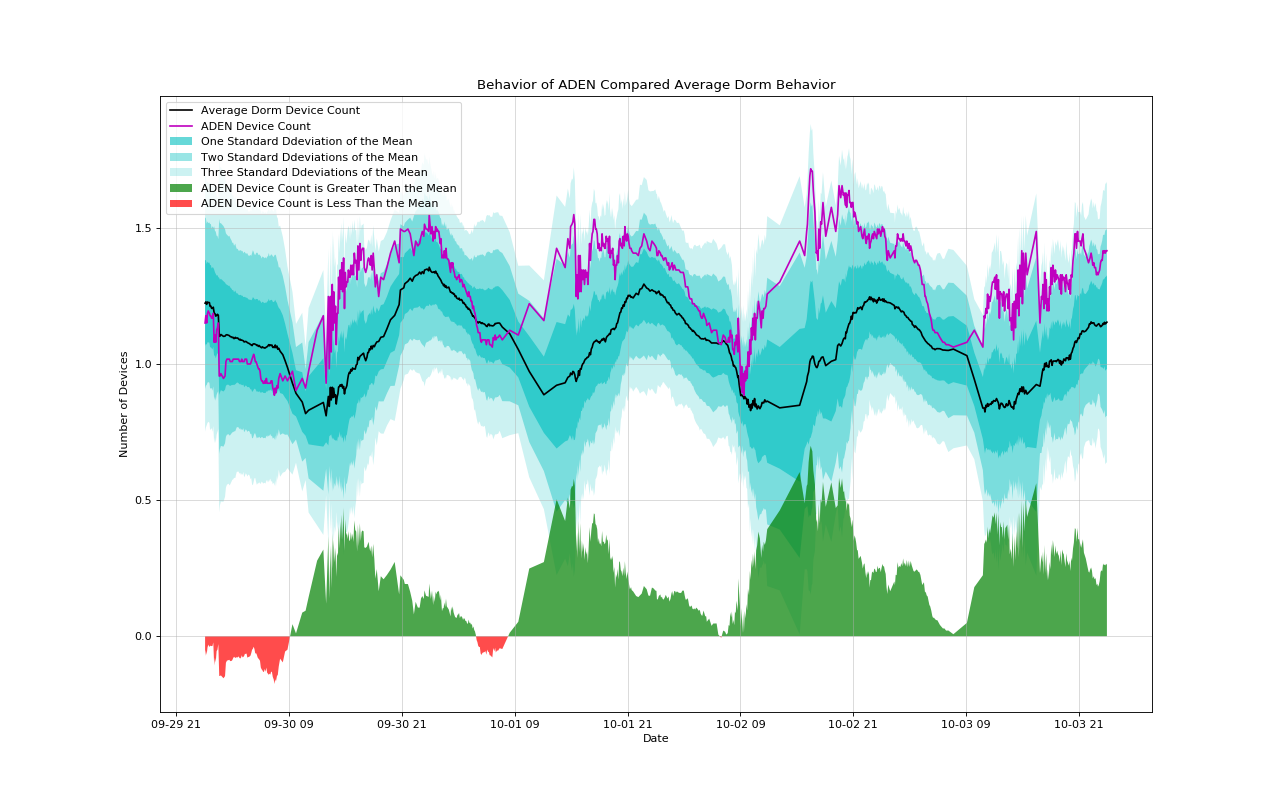}
    \caption{Aden Hall compared to the mean dorm behavior for a normal week in Fall 2019. For all mean dormitories images "Red areas" correspond to fewer devices than the mean and "Green areas" correspond to more devices than the mean.}
    \label{fig:ADENMean}
\end{figure}

Aden Hall, figure \ref{fig:ADENMean}, consistently has higher than average device density for a dorm. During peak usage, Aden can climb as high as 0.5 devices per person higher than the average, and will generally fall three standard deviations above the mean dorm. There are occasional times when the Aden device density drops below the mean; these often happen late at night between the hours of 11PM and 8AM indicating that students in this dorm sign off earlier than average.

\section{The COVID-19 Perturbation}

During Spring 2020, the rise and spread of COVID-19 resulted in a series of COVID related announcements, culminating in the campus asking students living in the dorm to leave. Using the notion of mean dorm behavior and using Fall 2019 as a baseline, we can see what effect that COVID had on the students and if the manner in which they left campus is considered typical or not. 

\begin{figure}[h!]
	\centering
    \includegraphics[width=0.45\textwidth]{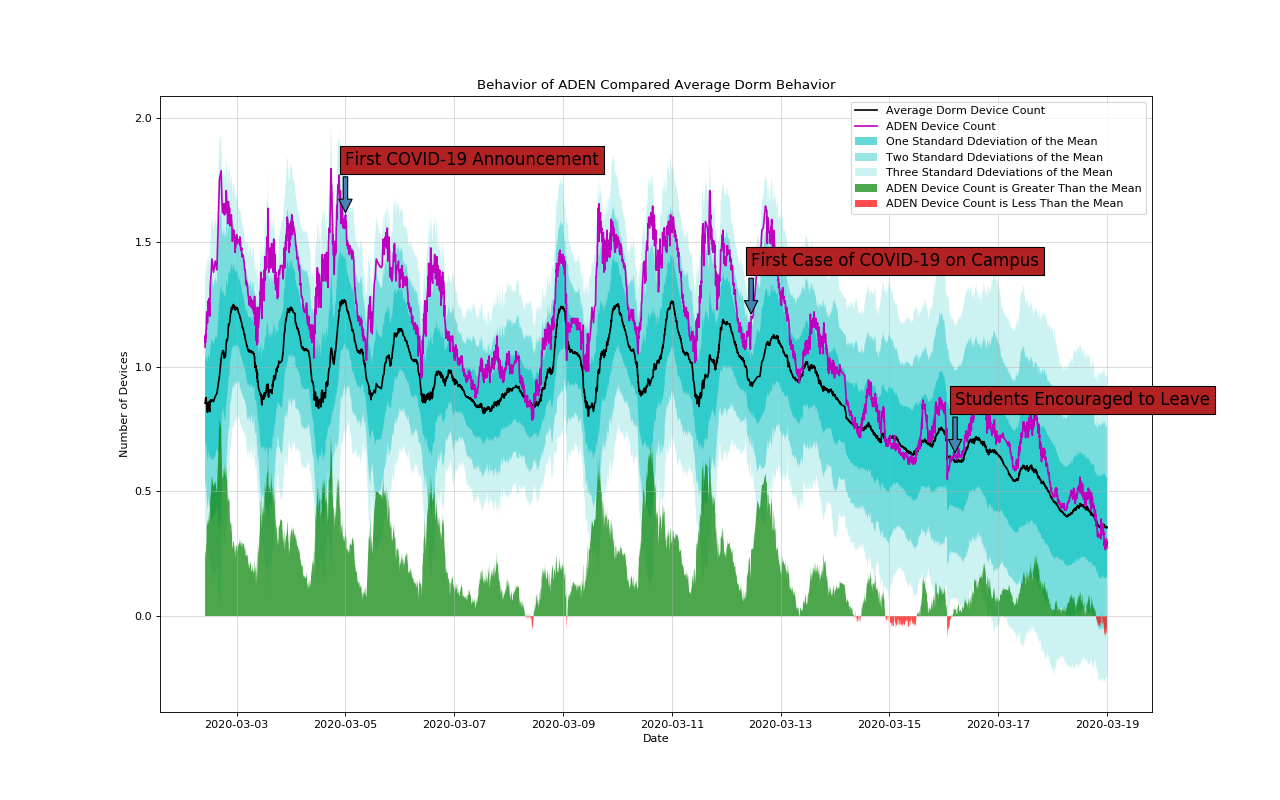}
    \caption{Aden Hall compared to the mean dorm behavior for the week students were encouraged to leave campus due to COVID-19, Spring 2020. Compare this image with raw data presented in "Fig. 1"}
    \label{fig:ADENCovid}
\end{figure}

Aden hall, figure \ref{fig:ADENCovid}, did not change behavior after the first official campus announcement of COVID-19. Device usage patterns and density remained constant until the first case appeared on campus. Almost immediately, students began to leave and device density in Aden hall swiftly dropped to within one deviation above the mean. In this dorm, students left at a rate very close to average; campus administration encouraging students to leave had no apparent effect on the rate in which they left. 

\subsection{The Rise of COVID-19 and The Timeline}

\begin{itemize}
    \item Thursday, March 5th: First announcement to CU Boulder staff, students and faculty of COVID-19, first two cases are confirmed in Colorado. 
    \item Monday, March 9th: Information begins to spread of the possibility of campus reaction to COVID-19 disease. 
    \item Tuesday, March 10th: Some official campus visits are cancelled. 
    \item Wednesday, March 11th: CU announces the movement to online classes, as well as cancelling official events over 150 people for the near future. 
    \item Thursday, March 12th: The first case of COVID-19 on campus is an employee who worked in the Center for Community Dining Center (C4C). 
    \item Friday, March 13th: In person classes cancelled. 
    \item Saturday, March 14th: Williams Village Recreation Center closes indefinitely, Main Campus Recreation Center operating on restricted hours. 
    \item Sunday, March 15th: Main Campus Recreation Center closes indefinitely, first CU Boulder student case confirmed. 
    \item Monday, March 16th: Sororities and Fraternities require that students move out by Friday, March 19.  Students are strongly encouraged to "return home."
    \item Tuesday, March 17th: CU Boulder 2020 commencement is cancelled. 
\end{itemize}{}

\section{Campus Response}
CU Boulder quickly responded to the pandemic, mandating that classes take an online or remote teaching approach soon after the recorded arrival of COVID-19 in Colorado. A day or two after the first recorded case arrived on campus, in person classes was cancelled. Subsequently, CU Boulder administration, over the the following week, transitioned to remote and online classes. It also encouraged students not to return to campus after the Spring 2020 break.

\subsection{A Naive Approximation and The Return That Never Was}
Our data shows that compared to a normal break, students generally left far sooner than expected. In some dorms, students began to move out as early as the week before break. Due to mandatory online classes, we would expect to see a higher device count in campus dorms, while in reality we saw quite the opposite.

A general observation is that trends in the data during the early parts of the semester do not vary significantly from week to week. As a result, we were able to quickly infer device behavior in a standard week from any other standard week in the year. With the COVID-19 perturbation to campus life, we used a quick but naive visual assessment by simply comparing one similar week to another to look for deviation in device behavior. We easily saw that students were leaving sooner than expected; we also quickly anticipated from looking at the data that, while not required at the time, students would stay away from campus after break. 

As spring break drew to a close, we monitored Wi-Fi activity in the dorms to look for any sign of returning students in anticipation of possibly needing to inform campus administration. Again, using a naive approach, we saw no signs of a "typical" return from a break, that is, when students returned to campus there was visible ramp up in device count. This was not noted for dates after spring break.

\begin{figure}[h!]
	\centering
    \includegraphics[width=0.45\textwidth]{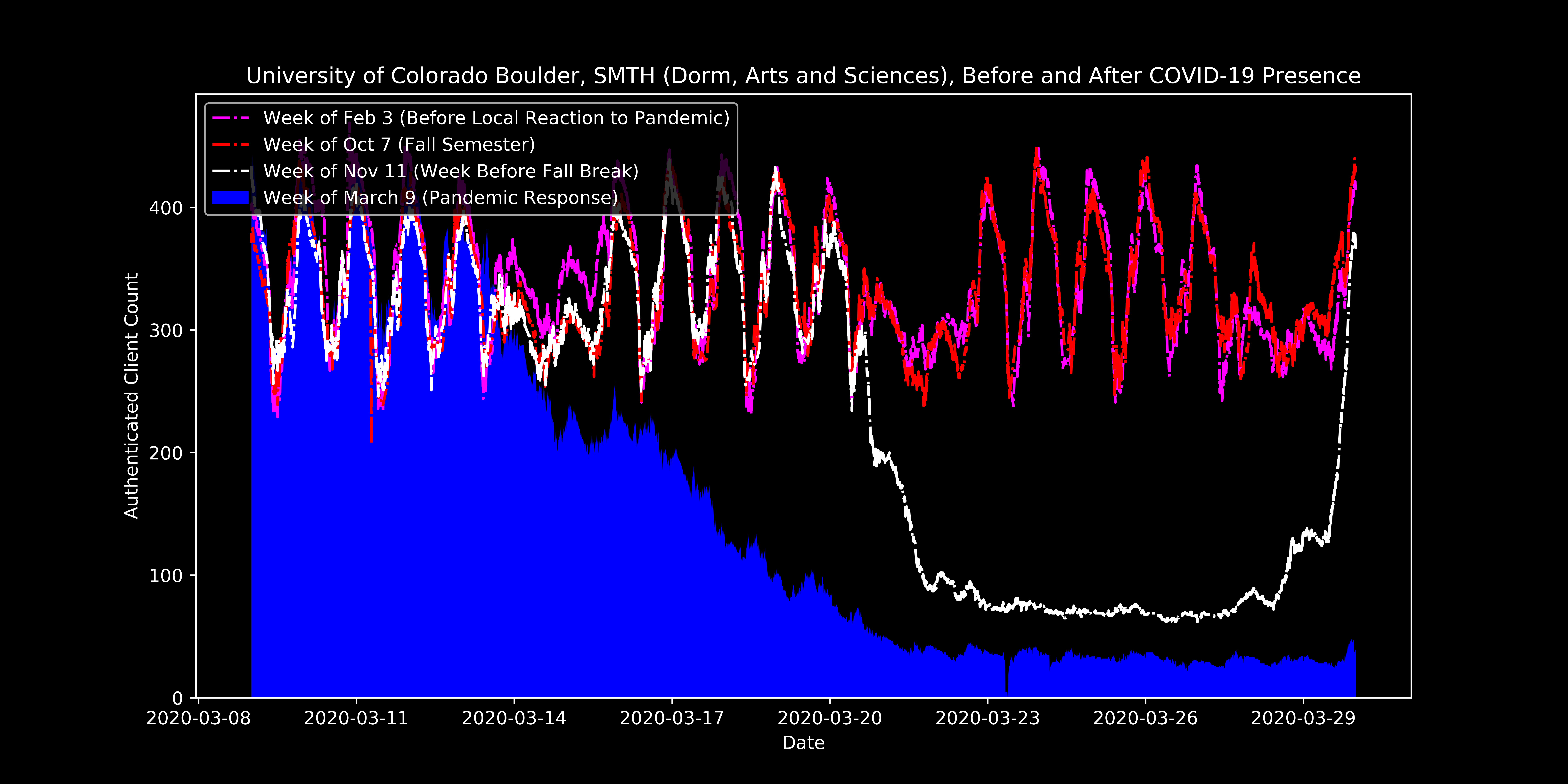}
    \caption{Smith Hall Wi-Fi device count for four weeks: October and November 2019 and February and March 2020 and compared to the mean dorm behavior for a normal week, Fall 2019. Analogue behavior is clearly visible for several weeks in the Fall 2019 and Spring 2020 as compared to indicated periods in the included box.}
    \label{fig:SMITHTypical}
\end{figure}

Figure \ref{fig:SMITHTypical} represents this naive approximation in the Smith resident hall. The white curve depicts behavior from Fall Break 2019 and how we might reasonably expect student managed devices to behave when students leave the dorm for an extended period of time. The magenta curve presents student device data in early parts of Spring 2020, before the pandemic, and the red line represents a standard week from the Fall 2019 semester. Finally, the blue area below the three upper curves depicts device response to the COVID-19 pandemic. We see that, in terms of general trends during "normal" weeks, one week's behavior is a relatively good predictor of another week's. The bundle of curves just mentioned, excluding the blue area, shows the usual ebb and flow of Wi-Fi activity at CUB. This makes the COVID-19 response look even more dramatic in comparison.

From these data, it is clear that students left campus roughly one week before they were expected to for spring break, but in a more gradual fashion. At the time, move out was not required, but encouraged, and it was unclear to students how long they would be away from their dorms. As a result, most devices had left by the time spring break would have started (note: the baseline of ~20 devices represents Wi-Fi enabled devices that are always connected to the network). As time progressed, the devices stayed away, and it became apparent that the return we would expect (the white line) was nowhere in sight.

As we proceed, a potentially useful scientific metaphor driving our research, because we are collecting access-point data from across campus, is an analogue to modern-day weather forecasting. More specifically, since such streaming data were collected at specific locations in the campus ecosystem and forecasts based on the spatial/temporal data were desired, we will view Singular Spectrum Analysis as being analogous to the statistical technique of empirical orthogonal functions (EOFs) used by L.F. Richardson and E. N. Lorenz in their quest for increasingly accurate weather predictions.

By summing the first 10 Eigen Triples in the decomposition figure 
we produce the rank-10 reconstruction that best approximates the initial streaming data and produce the image shown in the next figure.


\begin{figure}[h!]
    \centering
    \includegraphics[width=0.45\textwidth]{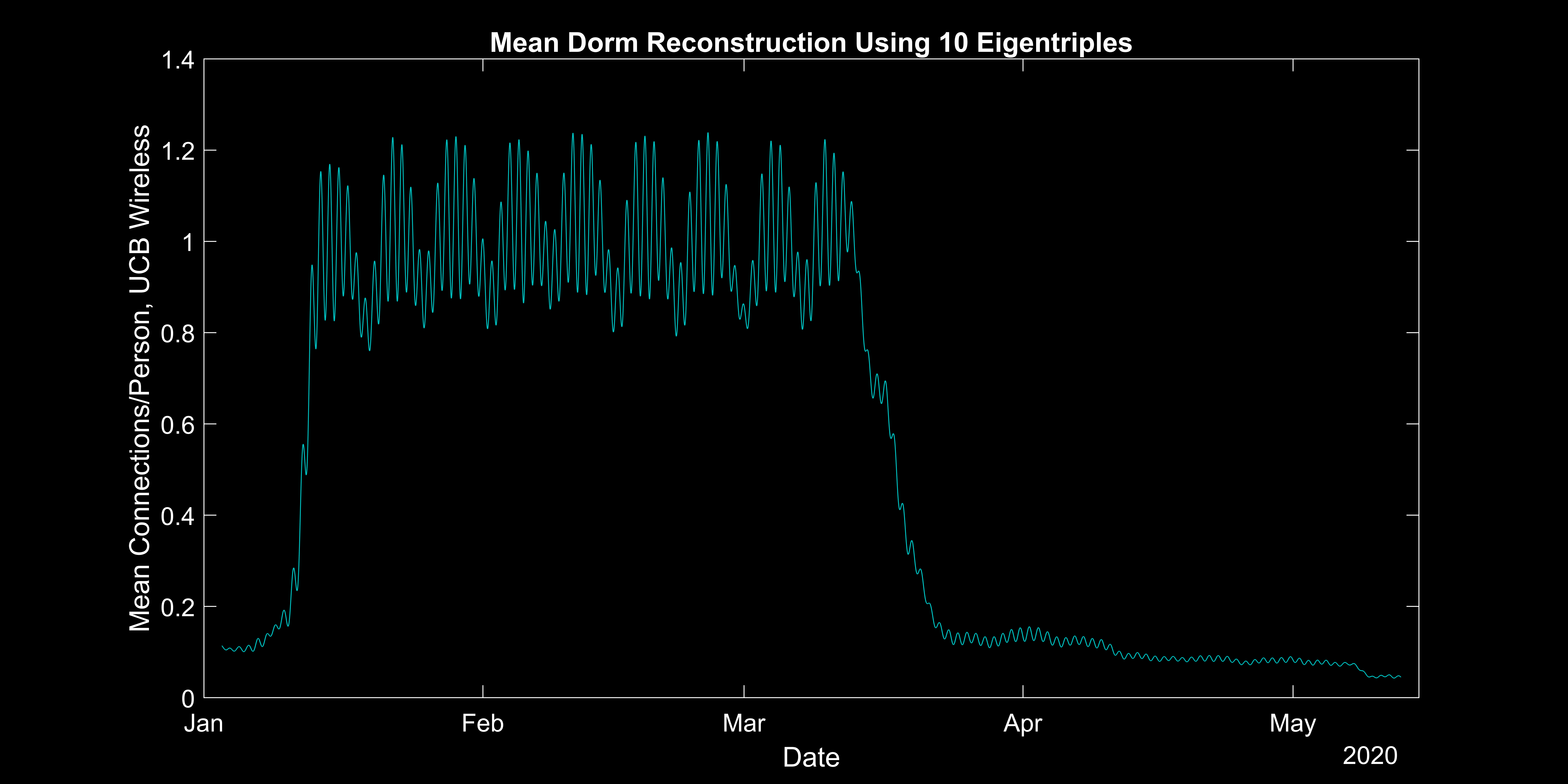}
    \caption{Normalized mean dorm device count reconstruction for the Spring 2020 semester using the first 10 Eigen Triples from our SSA decomposition. Dates used were January 3rd to May 13 th, all reconstructed values less than zero were considered to be zero.}
    \label{fig:MeanDormRecon}
\end{figure}

\begin{figure}[h!]
    \centering
    \includegraphics[width=0.45\textwidth]{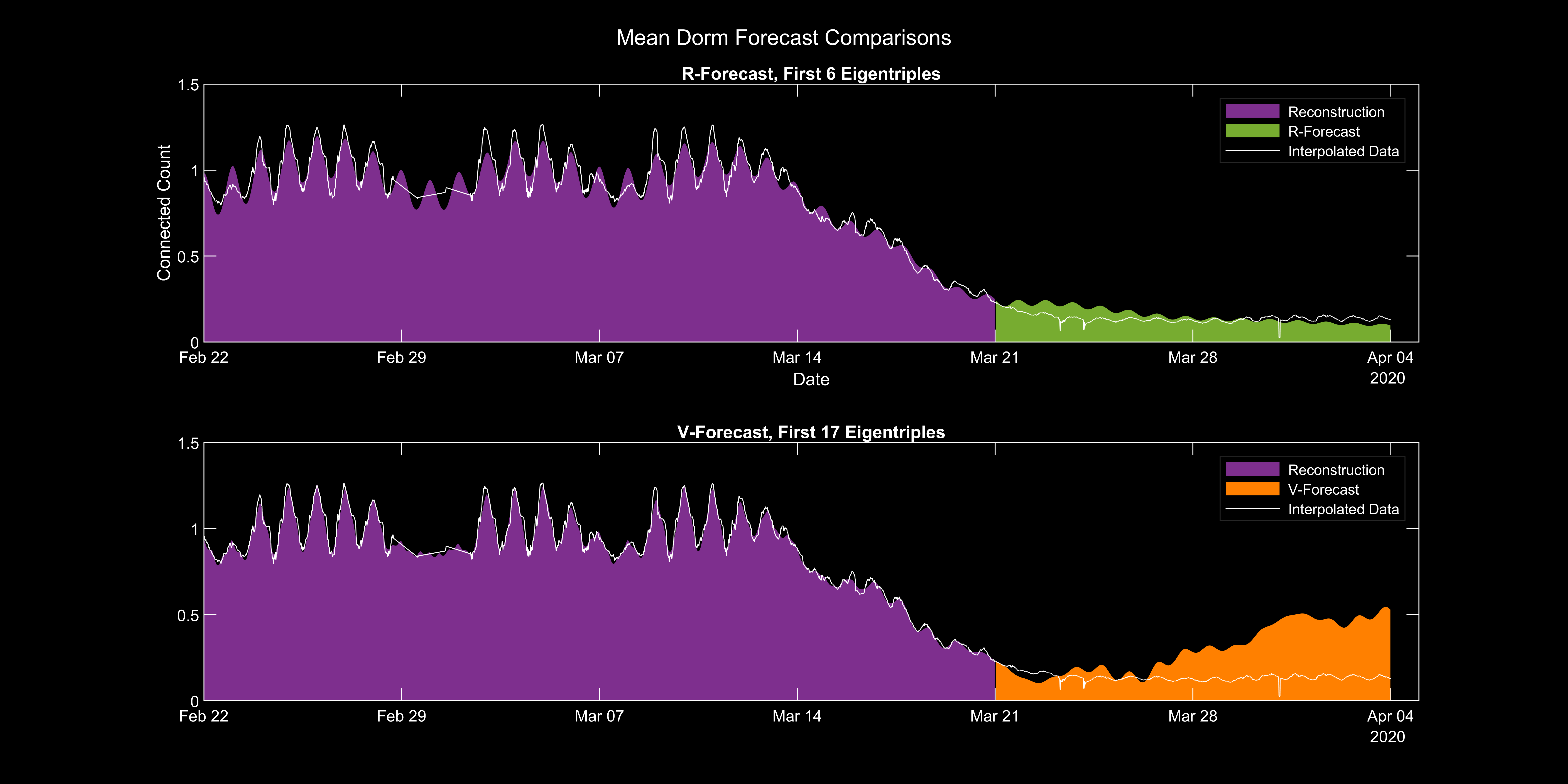}
    \caption{R-Forecast and V-Forecast using SSA analysis. Training data included portions of the COVID-19 Perturbation data. Dates used were March 21, 2020 to April 4, 2020. SSA R-forecasting strategy generated reasonable predictions while SSA V-Forecasting did not. Interpolation of the data, the white curve, was most reliably followed by R-Forecasting.}
    \label{fig:MeanDormRecon}
\end{figure}

The mean dorm is one crucial aspect of estimating populations on campus, since there must be a normalizing factor given a gross Wi-Fi connection count. After closer inspection, it is worth noting that peak device densities occur roughly at midnight while the lowest device densities occur roughly at midday. This is indicative of dispersal for classes during daytime hours and plays an important role for population estimation discussed further in later sections. The implication being that we must consider data during the nighttime when most students are inside their respective residence halls. 

\section{ Findings, Discussion, Concluding Remarks, and Future Work}

\smallskip
An initial goal of this research was to better understand the dynamics of the ebb and flow of aspects of campus life in the ecosystem on a major residential research university as reflected by Wi-Fi count dynamics. Our strategic and operational approach was to do this by working with CUB's Office of Information Technology (OIT) and specifically by collecting Wi-Fi usage data at scale. By data at scale, we mean that the volume of daily data exceeds multiple gigabytes, and therefore there is quite a lot of Wi-Fi connected device count data to process. This was an observable variable that was a proxy/surrogate for campus population dynamics.

This research’s point of view is that Wi-Fi at CUB not only presents a rich source of information, but also that its granularity could lend itself to differential data-driven prediction (DDDP) modeling, that is differential equation models generated from data. For example, at CUB Wi-Fi data sets contain 57,109,890 entries and with a point density of 1472.42857143 points/day. Further, and more importantly for this research's point of view, it offers a surrogate for the campus population.

Many new learning modalities and technologies demand network connectivity for functionality. Further, and because CU Boulder is a research university, more research devices are also demanding network access. Identifying such devices provides a baseline for activities, examples of these devices include vending machines in the AERO Space engineering building.

\smallskip

\subsection {Building Clusters} This research demonstrates that buildings in the CU Boulder ecosystem are of two types: dorms and "everything else." Dorms have nearly all contribution covered by the "first eigen triple"  (ET1), implying that they have a high Wi-Fi baseline of devices being used. Daily and weekly oscillations are discernible in the subsequent eigen triples, and each account for significantly less variation. However, both are roughly equal in their contribution to the original signal. The conclusions that we reached was that dorms have a high baseline of devices and exhibit approximately equal contributions due to daily and weekly variations. 

Other buildings in the ecosystem (C4C, Rec, Aero, ...) exhibit much less contribution by ET1 due to the extremely high daily variation and low baseline connectivity. Despite different numbers of Wi-Fi devices, each of these non-dorm buildings have the same structural frequency hierarchy. Explicitly, ETs 4 and 5 always seem to account for weekly oscillations that are independent of the building and ETs 2 and 3 are always a part of the daily oscillations, again, independent of the building. We interpret this as suggesting that these buildings have similar usage behavior in addition to daily oscillations being dominant over weekly oscillations. Further, facilities that do not house students have similar behavior even though the number of people inside can be significantly different between them. If we only considered ETs 1 through 5, each one of these buildings would be in clusters, i.e. have the same grouping. Finally, a surprise for this research was the number of devices, e.g. Vending machines, that routinely access Wi-Fi and add to a building's ecosystem.

\subsection {SSA Forecasting:}
Material on R-Forecast and V-Forecast using SSA analysis is presented in chapter 3 of \cite{ golyandina2001analysis}. As noted in \label{fig:MeanDormRecon} some training data included portions of the COVID-19 Perturbation data. In this case the SSA R-forecasting strategy generated reasonable predictions, while SSA V-Forecasting did not. We conjecture that it is likely that more can be done using other matrix  decompositions beyond  SSA, e.g. Non-Negative factorization is a good prospect.

\subsection {Technology, Cybersecurity, and Policy:} 
Wi-Fi at present is likely the most popular form of wireless technology that is adapted by IHEs to provide network services to students, faculty and staff. Wi-Fi officially launched at CU Boulder in September 1999 and has been evolving ever since with faster data rates and better security. Wi-Fi currently operates on unlicensed frequencies in the 2.4GHz and 5GHz band, making it possible for any enterprise to deploy a Wi-Fi network that enables its stakeholders to be mobile and to connect to the private and public networks. As a result, data density is increasing rapidly. However, legacy Wi-Fi installations are present and developing current maps of access points would be worth looking into as modeling and understanding based on data analytics becomes more valued.

As expected organizations governing standards and certification bodies , e.g. IEEE and the Wi-Fi Alliance, continue to improve Wi-Fi standards and ensure compatibility between Wi-Fi devices from over 800+ Wi-Fi product companies. Due to the popularity of Wi-Fi and the ability for Wi-Fi to positively impact GDP, the FCC is considering opening more spectrum that could leverage the Wi-Fi investment. This means that with the eventual roll out of 5G, spectrum sharing, and more, smart infrastructures will be developed and deployed. Understanding the impact on not only IHEs but societal infrastructure will become increasingly important. The interplay and connections are forcing the emergence of new languages, new analysis tools, and ultimately better forecasting of the 'weather' in such ecosystems, thus leading to the possibility of new ways of thinking about issues of Technology, Cybersecurity and Policy. 

\subsection {Future Work} 
Since IHEs are at the forefront of components of emerging trends in demand for connectivity, computation, connections, data flows, and innovations they are excellent environments to deploy passive and active sensors an harvest data streams. Melding the computational and mathematical tools that are emerging almost lend themselves to a near real time understanding of these vital ecosystem. The point of view of this research is that much can be learned about smart and connected environments through an examination of data streams. This research is looking forward to the "return" of faculty staff and students and using some of the tools mentioned here to extract further patterns of emergence in the IHE ecosystem.

\section{ Acknowledgments}

 Special thanks to OIT and the Campus Chief Security Officer for helping to make campus Wi-Fi data available for analysis and modeling. Many of the ideas and framing were developed because of colleagues in Technology, Cybersecurity, Policy program and from seminar presentations in that program. Special thanks to speakers: John Scrano of Zayo, Tom Kellerman of VMware, and Professor Kristine Kaminski of Wolf Law and Dr. Amie Stefanovic, and my colleague Dr. Kevin Gifford. Thanks also to Joshua Park for aiding in the editorial contributions to this article; Jackson Trust for several formative discussions and finally, thanks to unnamed colleagues for providing feedback on earlier versions of the manuscript.

\medskip

\section{Biography}

\begin{IEEEbiography}[{\includegraphics[width=1in,height=1.25in,clip,keepaspectratio]{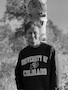}}]{First Author.} Jake McGrath is an MS. student at the University of Colorado Boulder he holds a BS in Aerospace Engineering Sciences with minors in Applied Mathematics and Leadership Studies with the President's Leadership Class form the University of Colorado Boulder.

\end{IEEEbiography}

\begin{IEEEbiography}[{\includegraphics[width=1in,height=1.25in,clip,keepaspectratio]{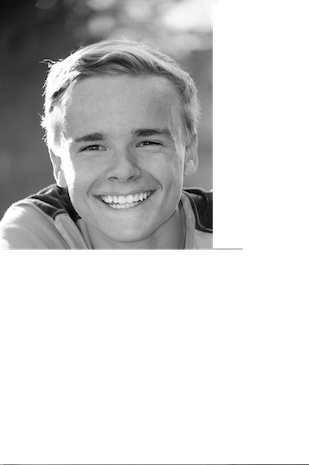}}]{Second Author,} 

Armen Davis is a Senior at the University of Colorado Boulder Majoring in Applied Mathematics and with a focus in the Atmospheric Sciences.

\end{IEEEbiography}

\begin{IEEEbiography}[{\includegraphics[width=1in,height=1.25in,clip,keepaspectratio]{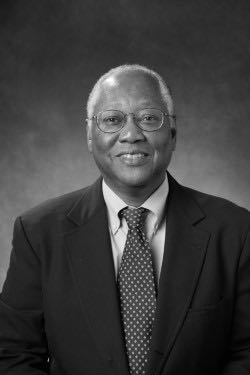}}]{Third Author}

Dr. James H. Curry,  Professor of Applied Mathematics and Computer Science, teaches in the Technology, Cybersecurity and Policy program at University of Colorado, Boulder. Curry is also a Member of  National Academies Board on Mathematical Sciences and Analytics (BMSA). His research interest include numerical methods, dynamical systems, linear algebra, data analytics. E-mail: james.h.curry@colorado.edu.
\label{fig}
\end{IEEEbiography}

\begin{IEEEbiography}[{\includegraphics[width=1in,height=1.25in,clip,keepaspectratio]{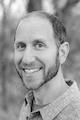}}]{Fourth Author.}
Orrie Gartner is the Director of Operations for the Office of Information Technology at at the University of Colorado Boulder. His focus is the application of IT to support and advance the mission of the University while overseeing daily operations and long term planning of reliable, high-performance, secure and scalable infrastructure.

\end{IEEEbiography}

 \begin{IEEEbiography}[{\includegraphics[width=1in,height=1.25in,clip,keepaspectratio]{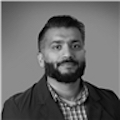}}]{Fifth Author, Jr.}
Glenn Rodrigues is a Senior Mobility and Wireless Architect at the University. Glenn leads Wi-Fi architecture design, configuration and support for one of the largest single site deployments of Wi-Fi in the state of Colorado. He is responsible for the vision, and driving an innovative and high ROI wireless strategy for the campus. Glenn introduced mapping customer mobility business needs and workflow into wireless technology solutions. 

\end{IEEEbiography}

\begin{IEEEbiography}[{\includegraphics[width=1in,height=1.25in,clip,keepaspectratio]{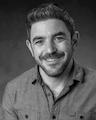}}]{Sixth Author.}

Dr. Seth Spielman is Associate Professor of Geography and Director of the University of Colorado Boulder office of Data Analytics. His research focus is spatial data, data analytics, modeling.
\end{IEEEbiography}

\begin{IEEEbiography}[{\includegraphics[width=1in,height=1.25in,clip,keepaspectratio]{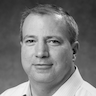}}]{Seventh Author.} Dr. Dan Massey is Professor of Computer Science and former Director of the Technology, Cybersecurity and Policy program.  His research focuses on cybersecurity for traditional systems, emerging cyber physical systems and the Internet of Things. Prior to joining CU, he was a Program Manager in the Department of Homeland Security, where his programs included denial of service defense, cyber physical system security for vehicles, medical devices and building controls.
\end{IEEEbiography}

\end{document}